# Dynamics of the lithium metal electrodeposition: Effects of a gas bubble

Shoutong Jin[1], Linming Zhou[1], Yongjun Wu[1, 2, *], Shang Zhu[3], Qilong Zhang[1], Hui Yang[1], Yuhui Huang[1], Zijian Hong[1, 2, *]

[1] School of Materials Science and Engineering, Zhejiang University, Hangzhou, Zhejiang 310027, China

[2] Cyrus Tang Center for Sensor Materials and Applications, State Key Laboratory of Silicon Materials, Zhejiang University, Hangzhou, Zhejiang 310027, China

[3] Department of Mechanical Engineering, Carnegie Mellon University, Pittsburgh, PA 16802, USA

Email: yongjunwu@zju.edu.cn (Y. W.); hongzijian100@zju.edu.cn (Z. H.)

**ABSTRACT**

**Rechargeable lithium metal batteries have been widely investigated recently, driven by the global trend for the electrification of transportation. Understanding the dynamics of lithium metal electrodeposition is crucial to design safe and reliable lithium metal anodes. In this study, we developed a grand potential-based phase-field model to investigate the effect of a static gas bubble, which forms due to the complicated internal side reactions, on the dynamics of the dendrite growth during electrodeposition. It is observed that with the presence of a gas bubble, the dendrite growth is largely accelerated, due to the accumulation of lithium ions on the far side of the bubble away from the anode surface, which could serve as an ion "reservoir" for the dendrite growth, leading to the bending/tilting of the lithium dendrites toward the bubble. Meanwhile, the effects of the bubble size and distance to the anode are further studied, demonstrating that the larger the bubble size and the closer to the anode, the longer the lithium dendrites grow. We hope this study could serve as an example to exploit the effect of extrinsic factors on the dendrite growth dynamics.**

There is a continuous trend for the complete electrification of transportation worldwide [1-3], to reduce carbon emissions and tackle environmental challenges such as global warming. One key requirement for the commercialization of electric vehicles is to increase the energy densities of lithium batteries. Lithium metal battery, the "holy grail" for batteries, remains one of the most promising solutions with high specific capacity and energy density [4-6]. Meanwhile, dendritic growth, the uncontrolled nonlinear growth of metal deposits, is one of the main issues that could result in battery failure and even severe safety issues for the lithium metal battery [7, 8]. Understanding the dendrite growth mechanism during electrodeposition is vital to designing the dendrite-free lithium metal battery.

Another non-negligible aspect for the degradation of lithium batteries is the gas generation, which occurs due to the complex interplay between the anode, cathode, and electrolyte [9-12]. For instance, it is discovered that the decomposition of the carbonate electrolytes could lead to the formation of multiple gases, including $H_2$, $CO$, $CH_4$, $C_2H_6$, and $CO_2$, etc. [10, 11], while the release of $O_2$ from the lattice oxygen is observed with ternary cathodes [12-14]. In particular, the carbonate electrolytes could react with chemical active lithium metal anode [15], forming mesoscopic (in the size of nm to mm) gas bubbles. These bubbles could either stick on the anode surface or release into the electrolyte, even resulting in the bulging of a soft pack. Previous experimental observation has shown that the formation of these gas bubbles could significantly affect the dynamics of the electrodeposition and the dendrite growth behavior for both half-cell and full-cell [16]. However, a comprehensive theoretical understanding of this phenomenon is still lacking.

On that front, phase-field methods have been widely adopted to understand the phase transitions, phase transformations, domain structures, and microstructure evolutions [17-19]. In particular, it has been widely employed to simulate the morphological evolution of lithium metal-based anodes [20-30]. These studies reveal the physical insights and critical parameters for the dendrite growth in lithium metal batteries, hoping to provide engineering solutions to realize dendrite-free lithium metal batteries.

In this study, we developed a grand potential-based phase-field model to investigate the effect of gas bubbles on the lithium metal electrodeposition kinetics and dendrite growth dynamics during electroplating. We discovered that the presence of static gas bubbles could significantly modulate the dendrite growth patterns, where those bubbles could act as a lithium-ion 'reservoir' on the semi-sphere external to the electrode. The local lithium enrichment gives rise to an additional driving force for the dendrite growth, while also leads to the bending of the dendrite. Further studies reveal that the larger the gas bubble, the faster the dendrite growth; whereas the shorter the bubble-anode distance, the quicker the dendrite nucleates. This study not only highlights the importance of the gas bubbles on the electrodeposition kinetics of lithium metal anode, but also provides an effective and insightful approach to reveal the non-negligible influence of extrinsic factors, e.g., bubbles, and impurities, on the performance of lithium metal batteries.

Following our previous works [20, 27], a grand potential-based phase-field model [31] is built where the non-conserved phase-field variable ξ is introduced as the phase order parameter, where ξ = 1 and 0 represent the pure electrode and electrolyte phases, respectively. At room temperature, the kinetic evolution of the order parameter ξ can be expressed as [20, 24]:

$$\frac{\partial \xi}{\partial t} = -L_\sigma (g'(\xi) - k\nabla^2 \xi) - L_\eta h'(\xi) \left\{ \exp\left[\frac{(1-\alpha)F\eta_a}{RT}\right] - \frac{c_{Li^+}}{c_0} \exp\left[\frac{-\alpha F\eta_a}{RT}\right] \right\} \quad (1)$$

where $L_\sigma$ and $L_\eta$ are the interfacial mobility and electrochemical reaction kinetic coefficient, respectively. $g(\xi)$ is the double-well function defined by: $g(\xi) = W\xi^2(1-\xi)^2$, $W$ is the switching barrier. $k$ is the gradient coefficient, which is related to the surface tension γ and interfacial width δ by $k = 6\gamma\delta$. $h(\xi)$ is the partition function. $\alpha$, $F$, $R$, $T$ and $c_0$ are the charge-transfer coefficient, Faraday constant, gas constant, temperature, and initial Li-ion molar ratio, respectively. $\eta_a$ is the activation overpotential, which is related to the applied overpotential $\varphi$, $\eta_a = \varphi - E^\theta$, where $E^\theta$ is the standard equilibrium half-cell potential. To model the gas bubble, an additional non-evolving order parameter $\zeta$ is introduced, where $\zeta=1$ represents the gas bubble. The lithium-ion concentration $c_{Li^+}$ is related to the chemical potential μ, order parameters ξ and $\zeta$, i.e.,

$$c_{Li^+} = \frac{\exp\left[\frac{(\mu-\varepsilon^l)}{RT}\right]}{1+\exp\left[\frac{(\mu-\varepsilon^l)}{RT}\right]} [1 - h(\xi)][1 - \zeta] \qquad (2)$$

where $\varepsilon^l$ is the difference in chemical potential for Lithium and the neutral component. The chemical potential can be obtained by solving the coupled electromigration equation, while the electric potential is calculated by solving the conduction equation [20]. The diffusivity inside the gas bubble is set as zero. The detailed derivation of the equations can be found in previous reports [20] and the Supporting Information.

The phase-field equations are solved using the open-source MOOSE framework [32, 33]. A two-dimensional mesh of 200 × 200 is employed, with each grid representing 1 μm. The simulation is performed at room temperature (300 K). The Newton method is used as the numerical tool, with the bdf2 scheme and single matrix preprocessing (SMP). Dirichlet boundary conditions are applied along the *X* dimension (perpendicular to the metal/electrolyte interface) for the phase-field variable ξ, chemical potential μ, and applied overpotential $\varphi$, while no flux boundary conditions are applied along the *Y* dimension (which is perpendicular to the nominal direction of the electric field). The left and right boundaries of the order parameter ξ are fixed at 1 (metal) and 0 (electrolyte), the chemical potential at both ends is fixed at 0 (fixed concentration for Li species on both sides), and the potential is set to the overpotential applied on the left boundary, and fixed to zero on the right boundary. A schematic diagram of the boundary conditions is given in **Fig. S1**. The primary simulation parameters before and after normalizations are listed in the **Table. S1**.

**Main**

The lithium electrodeposition process with and without a gas bubble under a relatively high applied overpotential (e.g., -0.22 V) is shown in **Fig. 1**. A 20 μm thick lithium metal anode is set as the initial configuration (**Fig. 1a**). The time-resolved lithium dendrite growth dynamics without the gas bubble are depicted in **Fig. 1(a)-1(d)**. It can be seen that after deposition for 150 s, the surface instability fluctuation shows up. During the consequent electrodeposition process, the small nuclei grow into several long filaments after 270 s, consistent with previous reports [20, 27]. The physical mechanism for the dendrite growth process has been attributed to the complex interplay between diffusion, electromigration, and electrodeposition [20]. Meanwhile, when a gas bubble with a radius of 20 μm is placed on the electrolyte 90 um away from the initial electrode, as depicted in **Fig. 1(e)-1(h)**, the dynamic behavior of dendrite growth changes drastically. It can be observed that the electrodeposition is severely hindered for the region shaded by the gas bubble, forming a concave-like structure. After electrodeposition for 200 s, with the presence of a gas bubble, the total number of dendrites is fewer, while the maximum dendrite length is much longer (**Fig. 1g**), which eventually becomes more prominent after 270 s (**Fig. 1h**). It is also interesting to note that in this case, the dendrites in the vicinity of the bubble tilt toward the bubble. The tilting of the dendrite could generate significant internal stress at the root of the dendrite, which may eventually lead to the breakdown of the dendrite, as has been observed in the previous experimental observation [16].

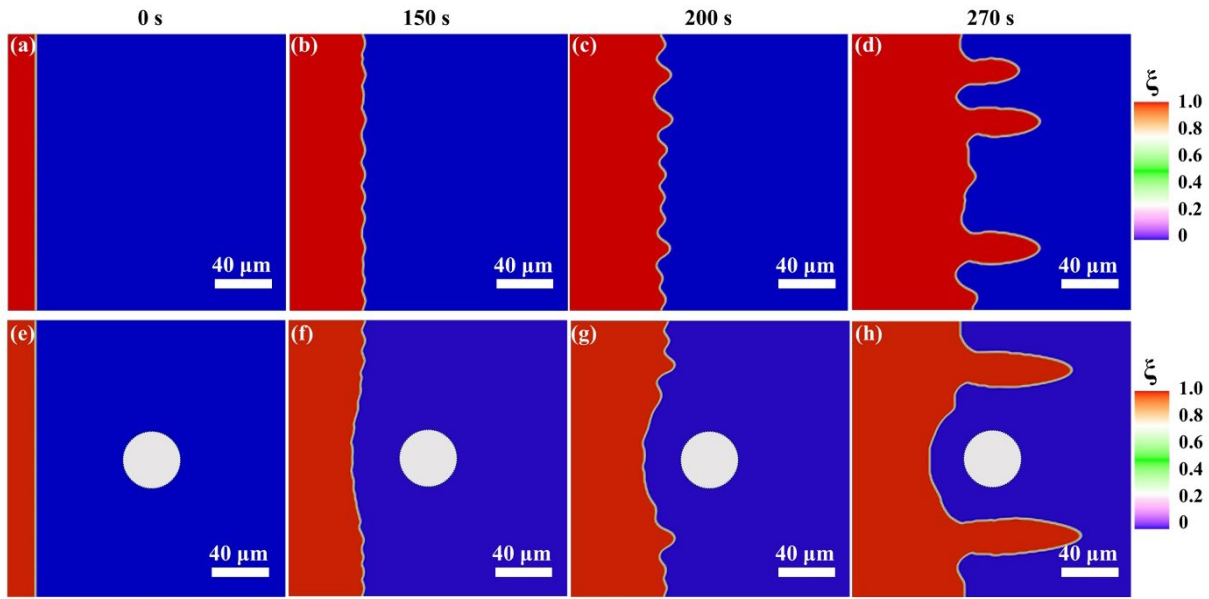

**Fig.1 Microstructure evolution for the lithium metal anode during electrodeposition (a-d) without and (e-h) with a gas bubble.** Electrode morphology without a bubble (a) The initial electrode morphology and after deposition for (b) 150 s, (c) 200 s, and (d) 270 s with an applied overpotential of -0.22 V. Electrode morphology with a bubble (e) The initial anode morphology with a gas bubble and the morphology and after deposition for (f) 150 s, (g) 200 s, and (h) 270 s with an applied overpotential of -0.22 V. The bubble radius is 20 μm, while the bubble center is 90 μm away from the initial electrode/electrolyte interface.

To further investigate the physical mechanism for the dendrite growth with gas bubbles, the Lithium-ion concentration distribution is plotted, as illustrated in **Fig. 2(a)-2(d)**. With the presence of a gas bubble, an enrichment of lithium-ion concentration can be observed on the right side of the bubble (away from the electrode), while the depletion of Lithium-ion concentration surrounding the left side of the bubble is close to the electrode (**Fig. 2a**). This can be understood since the bubble with zero lithium-ion diffusivity can severely hinder the lithium-ion transport, which causes local ion enrichment/depletion near/away from the lithium-ion source. After electrodeposition for 150 s, it can be discovered that a fan-shaped Lithium-ion concentration enrichment area is formed, showing the further increase of Lithium-ion concentration on the right side of the bubble (**Fig. 2b**). While between the bubble and the anode, the Lithium-ion concentration has been depleted. This further prevents the deposition of lithium metal beneath the bubble. Whereas after longer electrodeposition time, the ion enrichment near the bubble starts to decrease as the dendrites grow longer. This can be understood since an additional transport pathway has been built with the growth of the dendrite, where the far side of the bubble act as the additional Lithium-ion source for the dendrite growth (**Fig. 2c-2d**). The line plot of the horizontal and vertical concentration distributions through the bubble center is obtained, as shown in **Fig. 2(e)-2(f)**. It can be seen that as the deposition time increases, the concentration of the left side of the bubble (close to the anode surface) decreases monotonically, but on the right side of the bubble (far away from the anode), the lithium-ion concentration first increases and then decreases after 150 s (**Fig. 2e**), consistent with the 2-dimensional contour plots. While along the vertical direction, the Li-ion concentration decreases monotonically over time, showing the continuous transport of Li-ion through the two sides of the bubble.

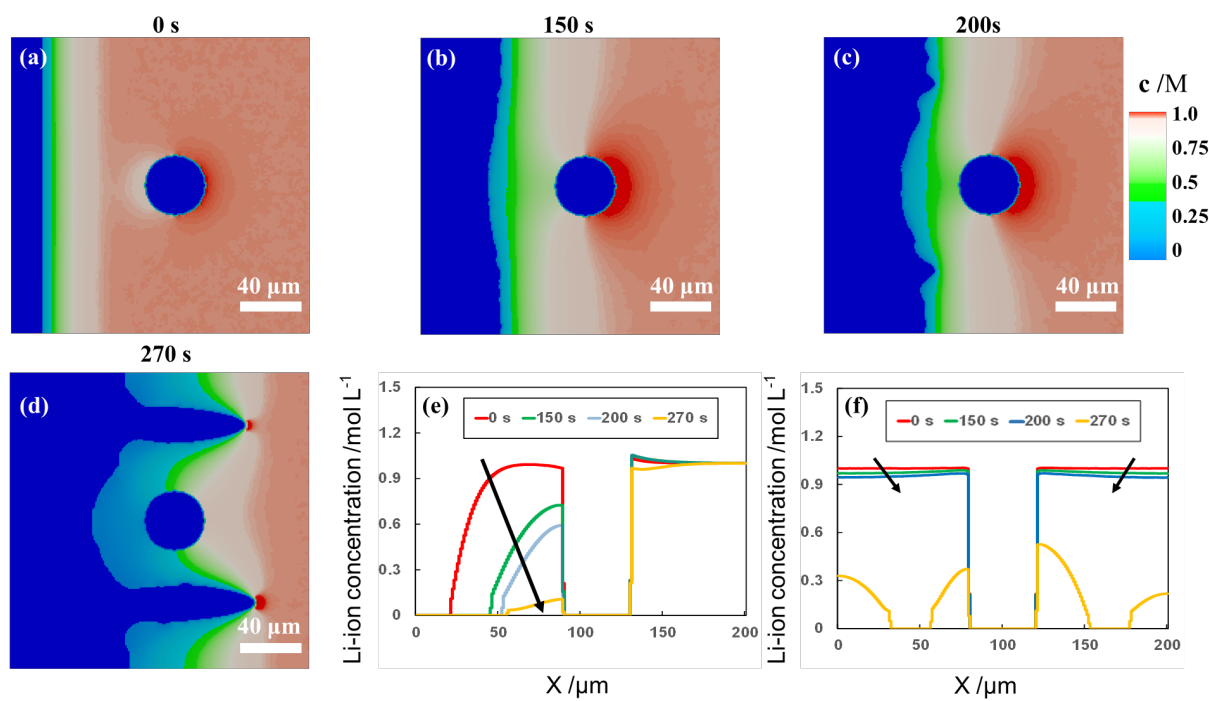

**Fig.2 Evolution of the Li-ion concentration with a gas bubble during electrodeposition.** Li-ion concentration distribution (a) at the beginning and after deposition for (b) 150 s, (c) 200 s, (d) 270 s at 0.22V.s. (e) The horizontal and (f) vertical Li-ion concentration distribution of the bubble center in 0 s,

150 s, 200 s, and 270 s, respectively at 0.22V.

In a short discussion, this reveals that the bubble could act as an ion reservoir during the initial deposition stage, which then provides an additional driving force for the consequent dendrite growth process. As a result, the growth direction of the dendrite could tilt toward the bubble region, leading to the bending of the dendrites. It should be noted here that this effect not only works for gas bubbles but any stuff that could potentially block the lithium-ion transport tunnel, including impurities in the electrolyte/separator, dead lithium, etc.

Having understood the physical insight of the dendrite growth kinetics with bubble, we proceed to investigate the effect of the bubble size on the electrodeposition. In this study, the center of the bubble is fixed while the size of the bubble changes from 4 μm to 20 μm (**Fig. 3**). As can be expected, when the size of the bubble is relatively small (e.g., <4 μm), it will not significantly affect the dendrite growth (**Fig. 3a**), after electrodeposition for 270 s, the shape and average length of the dendrites are similar to the case without a bubble (**Fig. 1d**). No apparent bending or tilting of the dendrites can be observed. Increasing the bubble size leads to longer dendrites after the same deposition time, with prominent tilting of the dendrites (**Fig. 3b-3c**). The Lithium-ion concentrations with different bubble sizes after electrodeposition for 270 s are plotted in **Fig. 3(d)-3(f).** It can be seen that as the bubble size increases, the ion enrichment area becomes more prominent on the far side of the bubble, while the shaded ion depletion region is also larger, resulting in the tilted growth of the lithium dendrite. This is confirmed by the line plot of the lithium-ion concentration, cutting through the center of the bubble along the horizontal axis (**Fig. 3g**), where the larger the bubble size, the higher the concentration difference between the two sides of the bubble. The dendrite length is further plotted against electrodeposition time for different bubble sizes (**Fig. 3h**), it can be discovered that the larger the bubble size, the shorter the dendrite nucleation time and the higher the dendrite growth velocity. This can be understood since with increasing the bubble size, the Lithium-ion "reservoir" becomes larger, providing a higher additional driving force for the lithium dendrite growth. This study indicates that while the large bubbles are detrimental to the electrodeposition, the smaller bubbles (with a radius <4 μm) have a moderate effect. It is suggested that practically, the control of the bubble size is key to effectively reducing the side effect of the gas bubble, which can be achieved either by reducing the gas reactions from the gas production side, as well as active perturbations during battery electrodeposition that could destabilize the giant bubbles.

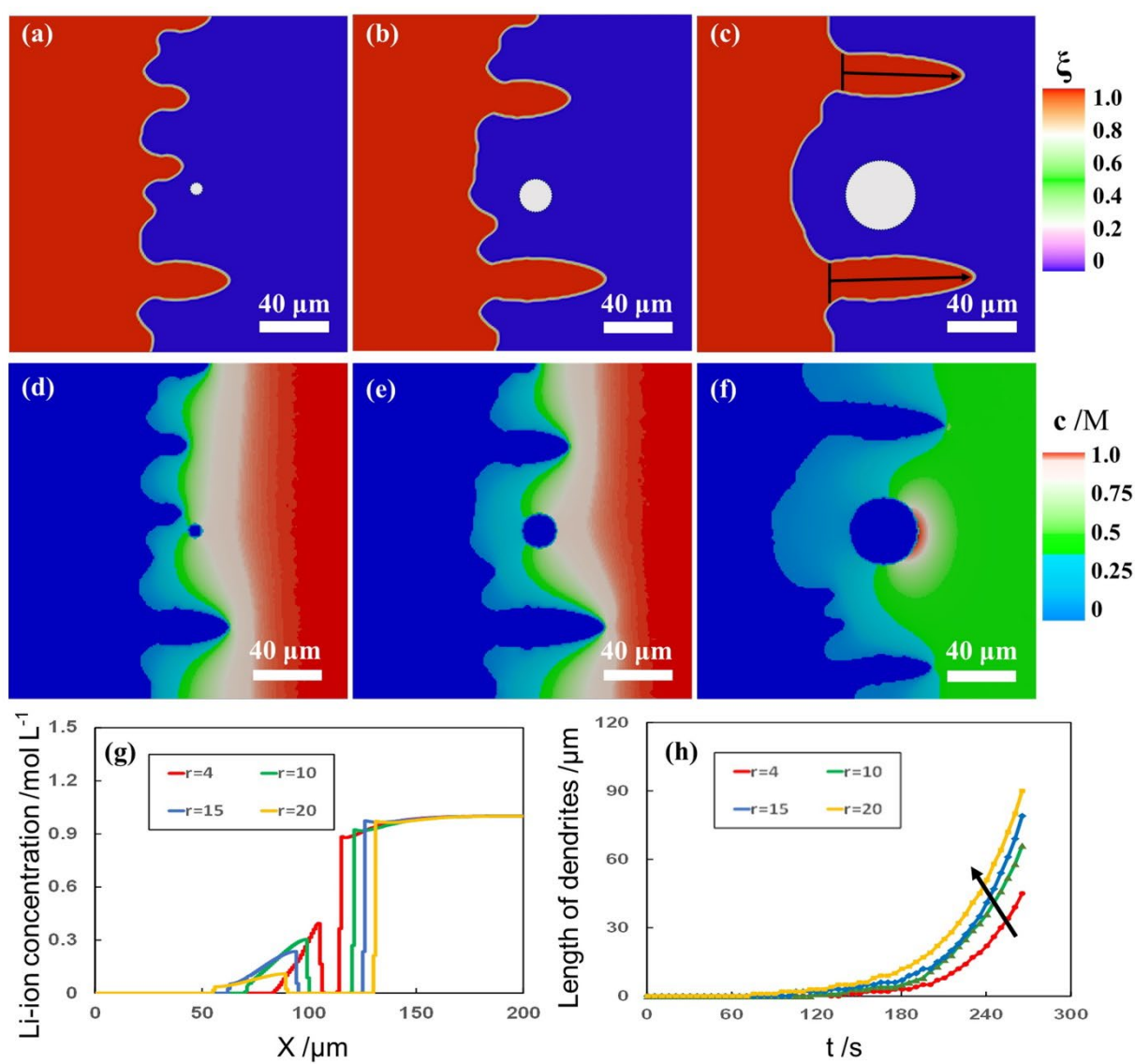

**Fig.3 Microstructure evolution for the lithium metal anode during electrodeposition with bubbles of varying sizes after deposition for 270 s.** (a-c) evolution and (d-f) Li-ion concentration. the

bubble has a radius of (a, d) 4 μm (b, e) 10 μm, and (c, f) 20 μm. (g) The horizontal Li-ion concentration distribution and (h) the dendrite length comparison of the bubble center with the bubble radius of 4 μm, 10 μm, 15 μm, and 20 μm.

The relative position of the bubble is also another critical aspect that could also affect the dendrite growth. The morphology of the Li metal anode after electrodeposition for 135 s with different bubble positions is given in **Fig. 4**. The center of the bubble is fixed at 60 μm, 80 μm, and 120 μm from the left edge of the initial anode position, while the size of the bubble is fixed at 20 μm. As shown in **Fig. 4(a)-4(c)**, the dendrite length is longer when the bubble is closer to the anode surface. This is also confirmed by the line plot of the dendrite length with respect to the electrodeposition time for different bubble-anode distances (**Fig. 4d**). It can be observed that when the distance is 60 μm, the dendrite nucleates near the two edges of the bubble on the anode surface, which grow into two long filaments that tilt towards the bubble, and after surpassing the bubble region, the growth direction becomes straight forward (**Fig. 4a**). The initial degree of tilting for the dendrites decreases with increasing bubble-anode distance. To understand the mechanism behind the dendrite growth rate and bubble-anode direction, the line plots of the ion concentration after different deposition time (10 s, 70 s, and 140 s) is shown in **Fig. 4(e)-4(g)**. It can be observed that while the concentration on the right side of the bubble is similar, the concentration decay on the left side of the bubble is more severe for the case when the bubble anode distance is shorter. After 140 s, the Lithium-ion on the left side of the bubble almost drained for the case when the bubble is very close to the anode **(Fig. 4g)**. The significant differences in Lithium-ion concentration between the bubble-shaded and free surface could result in the variation of the local plating rate, giving rise to preferred nucleation sites near the edge of the region shaded by the bubble. This effect is less severe where the bubble is further away from the anode surface. In a short discussion, the bubbles that stick on the anode surface are detrimental to the reversibility of the plating/stripping, while the bubbles in the middle of the electrolyte have a minimal effect. Reducing the bubble adhesion on the anode surface is another effective way to suppress dendrite nucleation and growth. This can be achieved by adding defoamers on the electrolytes, etc.

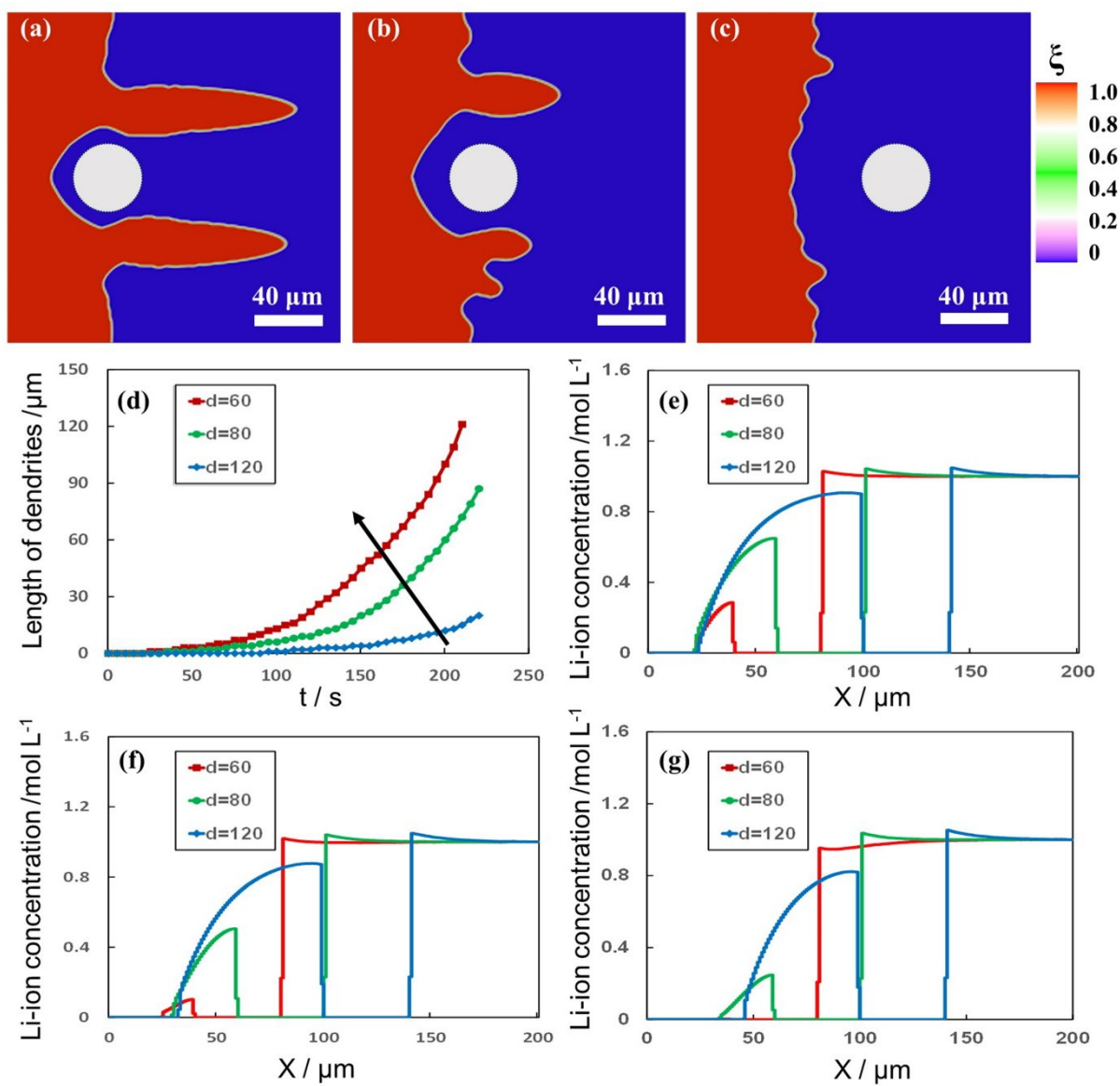

**Fig.4 Microstructure evolution for the lithium metal anode during electrodeposition with bubbles at varying positions, after deposition for 214 s.** Morphological evolution with a bubble at a distance of (a) 60 μm, (b) 80 μm, and (c) 120 μm. (d) The length of dendrites comparison of the bubble in the horizontal position. Horizontal concentration distribution of bubbles in the positions of 60 μm, 80 μm, and 120 μm after (e) 10 s, (f) 70 s, and (g) 140 s.

In conclusion, we have developed a grand potential-based phase-field model to investigate the effect of a static gas bubble on the lithium dendrite growth dynamics. It is revealed that the presence of the gas bubble causes significant dendrite growth, where the bubble side external to the electrode acts as a lithium-ion 'reservoir'. It provides an additional driving force for the dendrite growth, leading to the tilting of the dendrite towards the bubble. While the drain of the lithium-ion on the region shaded by the bubble could lead to uneven growth on the metal anode that could act as the preferred nucleation sites at the edge of the bubble. It is further revealed that a larger and closer-to-anode bubble triggers the growth of longer dendrites with shorter nucleation time. This finding provides promising engineering solutions for stabilizing lithium metal batteries by decreasing the bubble size and removing those bubbles near surfaces. We hope this study could spur further theoretical/experimental efforts in the complicated interactions between dendrite growth and extrinsic factors such as gas bubbles.

**Acknowledge**

A start-up grant from Zhejiang University is acknowledged (Z. H.). This work is supported by the Fundamental Research Funds for the Central Universities No. 2021FZZX001-08 (Z. H.). The simulations are performed on the MoFang III cluster, Shanghai Supercomputing Center (SSC).

**Notes**

The authors declare no competing interests.

# Supporting information for:

# Dynamics of the lithium metal electrodeposition: Effects of a gas bubble

Shoutong Jin[1], Linming Zhou[1], Yongjun Wu[1,2,*], Shang Zhu[3], Qilong Zhang[1], Hui Yang[1],

Yuhui Huang[1], Zijian Hong[1,2,*]

[1] School of Materials Science and Engineering, Zhejiang University, Hangzhou, Zhejiang 310027, China

[2] Cyrus Tang Center for Sensor Materials and Applications, State Key Laboratory of Silicon Materials, Zhejiang University, Hangzhou, Zhejiang 310027, China

[3] Department of Mechanical Engineering, Carnegie Mellon University, Pittsburgh, PA 16802, USA

Email: yongjunwu@zju.edu.cn (Y. W.); hongzijian100@zju.edu.cn (Z. H.)

*Phase-field model.*

Following the previous works [1-4], a grand-potential based phase-field model is built, with both the phase-field variable ξ and the chemical potential μ selected as the primary order parameters. The evolution of the phase-field variable ξ is governed by the Bulter-Volmer kinetics:

$$\frac{\partial \xi}{\partial t} = -L_\sigma(g'(\xi) - k\nabla^2\xi) - L_\eta h'(\xi)\left\{\exp\left[\frac{0.5F\eta_a}{RT}\right] - \frac{c_{Li^+}}{c_0}\exp\left[\frac{-0.5F\eta_a}{RT}\right]\right\} \quad (1)$$

Where $t$, $L_\sigma$, $L_\eta$ are the evolution time step, interfacial mobility, and electrochemical reaction kinetic coefficient, respectively. $g(\xi)$ and $h(\xi)$ are the double-well functions and partition function. $k$, F, R, T and $c_0$ are the gradient coefficient, Faraday constant, gas constant, temperature, and initial Li-ion molar ratio, respective. $\eta_a$ is the activation overpotential, which is related to the applied overpotential $\varphi$ for a half-cell, $\eta_a = \varphi - E^\theta$, where $E^\theta$ is the standard equilibrium half-cell potential. The local lithium-ion concentration $c_{Li^+}$ is related to the chemical potential μ, order parameters ξ and bubble order parameter ζ (which is defined as 1 for the bubble region and 0 for the other area), e.g.,

$$c_{Li^+} = \frac{\exp\left[\frac{(\mu-\varepsilon^l)}{RT}\right]}{1+\exp\left[\frac{(\mu-\varepsilon^l)}{RT}\right]}[1-h(\xi)][1-\zeta] \qquad (2)$$

where $\varepsilon^l$ is the difference in the equilibrium chemical potential for lithium-ion and the neutral component in the electrolyte. The chemical potential $\mu$ can be obtained by solving the coupled transport and electromigration equation [1]:

$$\chi\frac{\partial\mu}{\partial t} = \nabla\cdot\frac{Dc_{Li^+}}{RT}[\nabla\mu + nF\nabla\varphi] - \frac{\partial h(\xi)}{\partial t}[c^s\frac{C_m^s}{C_m^l} - c^l] \qquad (3)$$

The local diffusivity D is calculated by: $D = D_0[1-h(\xi)][1-\zeta]$, where $D_0$ is the equilibrium diffusivity for Li-ion in the 1 M LiPF$_6$ electrolyte. $C_m^s$ and $C_m^l$ are the site density of electrode and electrolyte, respectively.

The susceptibility $\chi$ can be expressed by:

$$\chi = \frac{\partial c^l}{\partial\mu}[1-h(\xi)] + \frac{\partial c^s}{\partial\mu}h(\xi)\frac{C_m^s}{C_m^l} \qquad (4)$$

The molar ratio for the liquid and solid phases, $c^l$ and $c^s$ can be obtained by:

$$c^{l,s} = \frac{\exp\left[\frac{(\mu-\varepsilon^{l,s})}{RT}\right]}{1+\exp\left[\frac{(\mu-\varepsilon^{l,s})}{RT}\right]} \qquad (5)$$

where $\varepsilon^{l,s} = \mu_0^{l,s} - \mu_N^{l,s}$ is the difference of the chemical potential for lithium and neutral component in the liquid and solid phases. The local electric potential $\varphi$ can be obtained by solving the following conduction equation:

$$\nabla\sigma\nabla\varphi = nFC_m^s\frac{\partial\xi}{\partial t} \qquad (6)$$

The effective conductivity is related to the order parameters $\xi$ and $\zeta$,

$$\sigma = [\sigma^s h(\xi) + \sigma^l(1-h(\xi))](1-\zeta) + \sigma^g\zeta \qquad (7)$$

$\sigma^s$, $\sigma^l$, and $\sigma^g$ are the conductivity for the lithium metal, electrolyte, and bubble, respectively.

*Boundary conditions.*

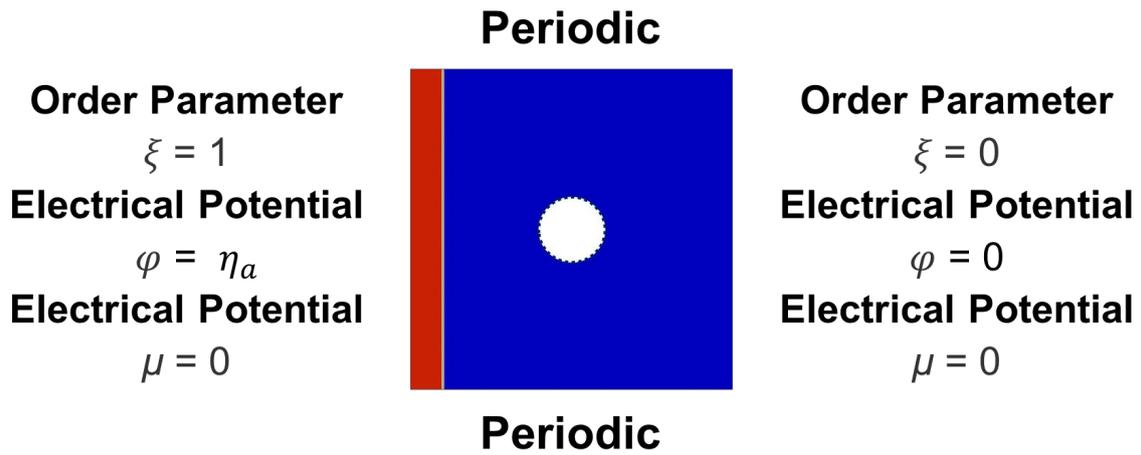

**Figure S1: Boundary conditions for the phase-field model.**

**Parameters used in the simulation.** In this study, a standard electrolyte with 1 M LiPF$_6$ dissolved in EC/DMC (1:1 volume ratio) solutions is assumed. A 2-dimensional simulation mesh of 200×200 is used, with each grid representing 1 μm. To simulate the initial local fluctuations of $\xi$ in the system, a Langevin noise level of 0.04 was added. All the parameters used in the current phase-field model are given in the following table:

Table S1: Phase-field simulation parameters

| Variable name | Symbol | Real value | Normalized value | Source |
|---|---|---|---|---|
| Interfacial mobility | $L_\sigma$ | $2.5 \times 10^{-6}\, m^3/(J \times s)$ | 6.25 | [5] |
| Limiting current | $i_0$ | 3 mA/cm$^2$ | 30 | [6] |
| Kinetic coefficient | $L_\eta$ | 0.1/s | 0.1 | Computed |
| Electrons transferred | $n$ | 1 | 1 | - |
| Surface tension | $\gamma$ | 0.556 J/m$^2$ | 0.22 | [7] |
| Interface thickness | $\delta$ | 1 $\mu$m | 1 | Estimated |
| Transfer coefficient | $\alpha$ | 0.5 | 0.5 | [5] |
| Site density of electrode | $C_m^s$ | $7.64 \times 10^4$ mol/m$^3$ | 76.4 | [5] |
| Site density of electrolyte | $C_m^l$ | $1.44 \times 10^4$ mol/m$^3$ | 14.4 | Computed |
| Conductivity of electrode | $\sigma^s$ | $10^7$ S/m | $10^7$ | [5] |
| Conductivity of electrolyte | $\sigma^l$ | 1.19 S/m | 1.19 | [8] |
| Conductivity of bubble | $\sigma^g$ | 0 S/m | 0 | - |
| Diffusivity of Lithium ion | D | $3.197 \times 10^{-10}$ m$^2$/s | 319.7 | [8] |

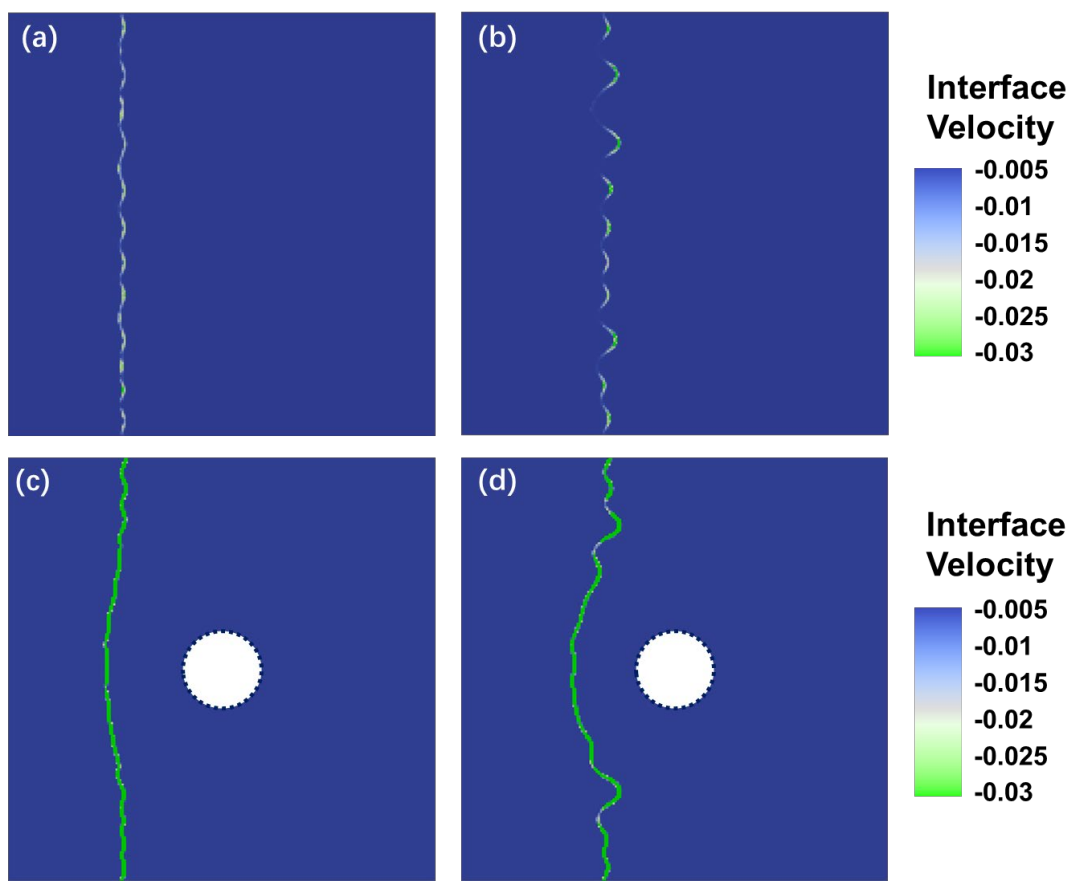

**Figure S2: The interface velocity $g$ at the lithium metal anode during electrodeposition** after 100 s and 135 s without (a-b) and with (c-d) a gas bubble under an applied overpotential of -0.22 V. The bubble center is 90 μm away from the initial electrode/electrolyte interface.

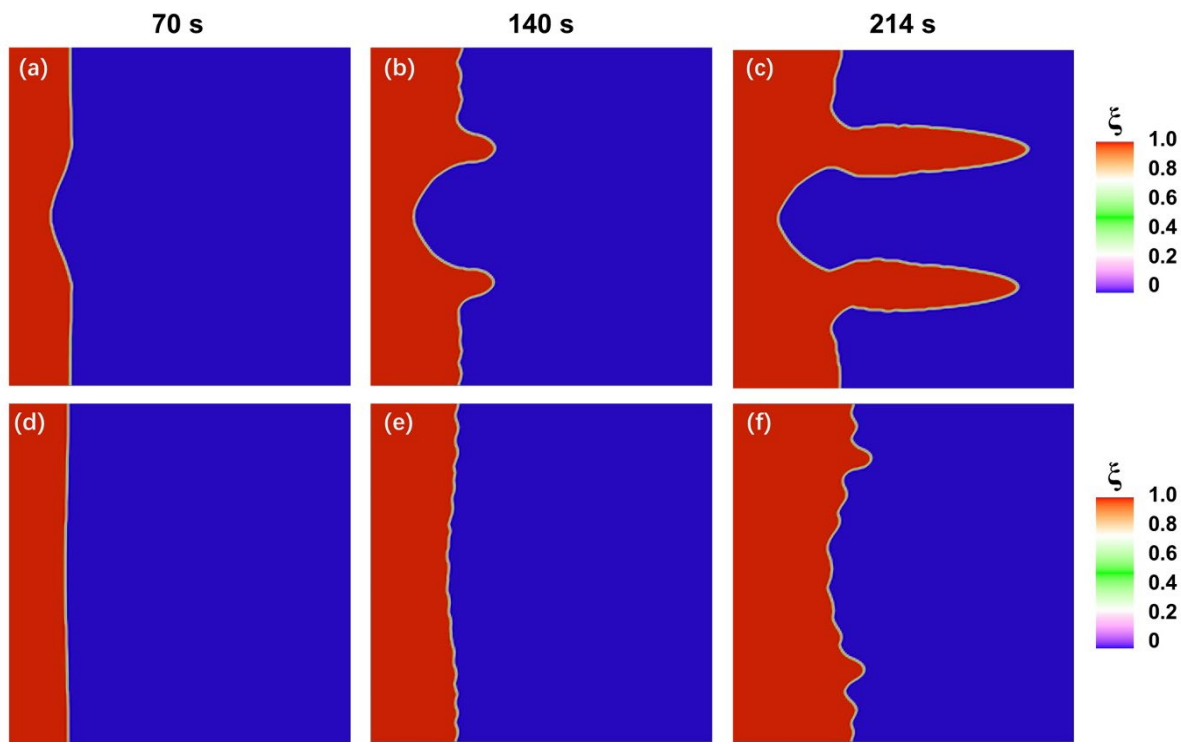

**Figure S3: Microstructure evolution for the lithium metal anode during electrodeposition**

**with a gas bubble at varying positions, after deposition for 70 s, 140 s and 214 s.** The bubble center is (a-c) 60 μm, and (d-f) 120 μm away from the initial electrode/electrolyte interface.

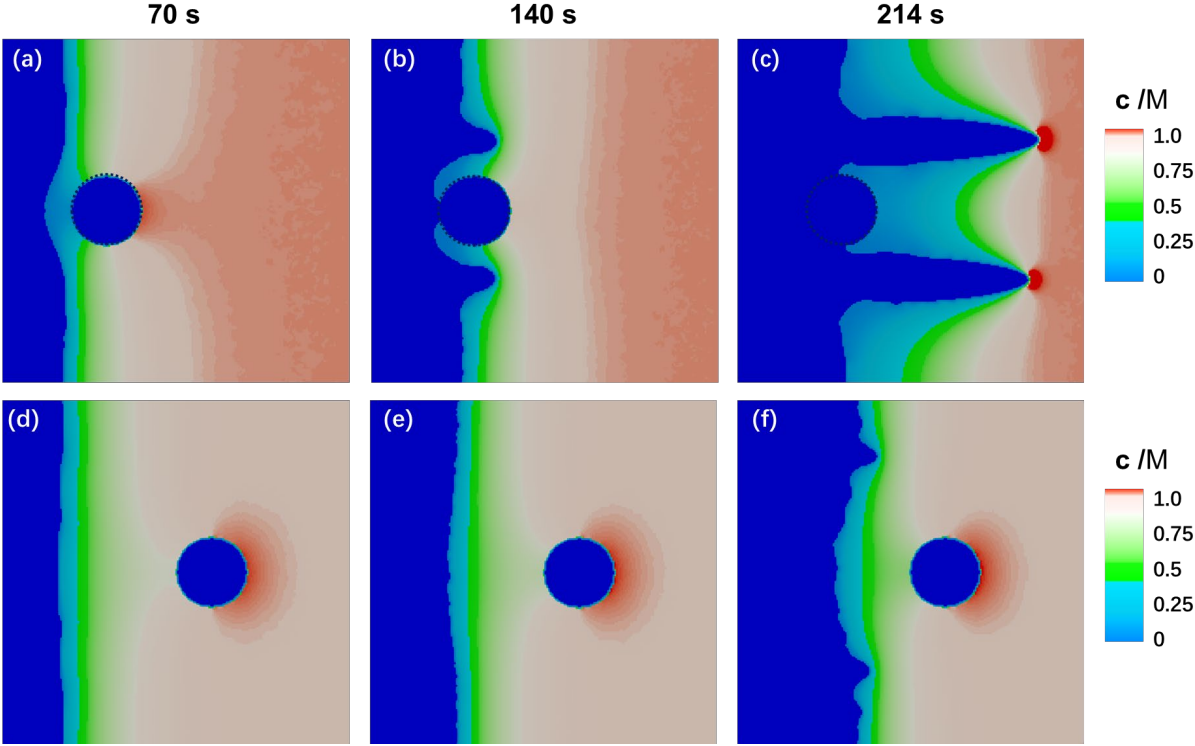

Figure S4: Spatial distribution of the Lithium-ion concentration, corresponding to each stage in Fig. S3.

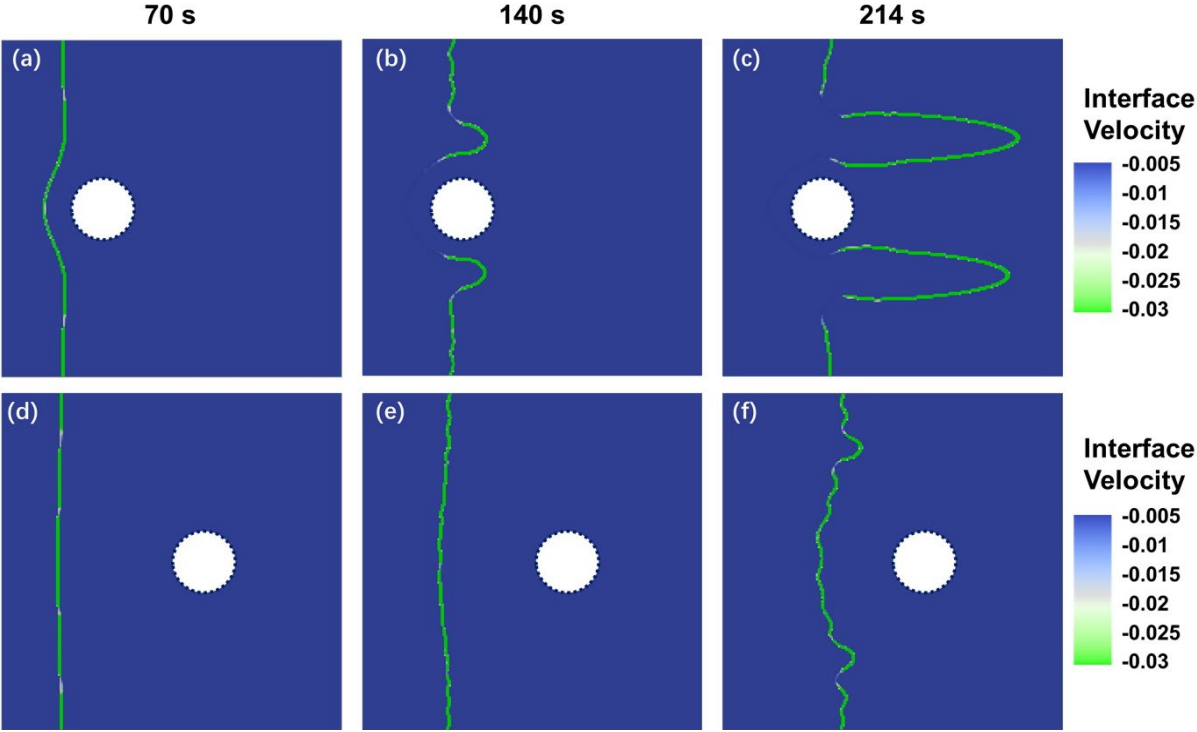

Figure S5: Time evolution of the interface velocity g, corresponding to each stage in Fig. S3.